\documentclass[aip,jcp,amsmath,amssymb,reprint]{revtex4-1}
\usepackage{hyperref}
\pdfoutput=1
\usepackage{graphicx}
\usepackage{amssymb}
\usepackage{amsmath}
\usepackage{blkarray}
\usepackage{multirow}
\usepackage{natbib}
\usepackage{epstopdf}
\usepackage{float}
\usepackage{breqn}
\usepackage[normalem]{ulem}
\usepackage{graphicx}
\usepackage{dcolumn}
\usepackage{bm}
\usepackage{t1enc} 
\newcommand{\dH}{\text{\normalfont\DH}}
\usepackage{tabu} 

\usepackage{mathrsfs}

\begin{document}

\title{Bridging the gap between mesoscopic and molecular models of solid/liquid interfaces out-of-equilibrium}
\author{Aaron R. Finney}%
\affiliation{Thomas Young Centre and Department of Chemical Engineering - University College London - London WC1E 7JE UK.}%
\author{Matteo Salvalaglio$^*$}%
\affiliation{Thomas Young Centre and Department of Chemical Engineering - University College London - London WC1E 7JE UK.}%
\email{m.salvalaglio@ucl.ac.uk}
\date{\today}
\begin{abstract}
Solid/liquid interfaces control various processes of technological relevance in the process industry and many fundamental physicochemical phenomena. 
This work examines the link between the atomistic description of mass transfer at solid/liquid interfaces out of equilibrium and the constitutive mass transfer equations typically used to model these processes at the mesoscale. In our analysis we discuss the microscopic inconsistencies apparent in simplified models of mass transfer whenever non-idealities dominate the liquid phase in the proximity of solid/liquid interfaces. Using C$\mu$MD - a molecular simulation technique to investigate out-of-equilibrium, concentration-driven processes with pseudo open-boundary conditions - 
we outline a strategy to capture and quantify non-idealities induced by specific interactions between the solid surface and molecules in the fluid phase.
We demonstrate our approach by studying electrolyte solutions in technologically important, multi-component systems in which introducing a solid/liquid interface induces substantial deviations in the composition and electrochemical properties compared to the fluid phase bulk. 
To this aim, we analyse NaCl(aq) solutions in contact with \emph{i}) the graphite basal surface and \emph{ii}) \{100\}) NaCl(s) crystalline surfaces. We uncover the tendency of the sodium cation to preferentially adsorb at the surfaces considered, which induces local violations of electroneutrality that leads to the emergence of an electric potential in the fluid phase at the solid/liquid interface. 
\end{abstract}

\maketitle

\section{Introduction}
\noindent Solid/liquid interfaces play a central role in a range of processes including catalysis, \cite{iwasawa_chemical_1987} crystal growth \cite{pimpinelli_physics_1998} and filtration \cite{burchell_carbon_1999}. In order to control these processes for e.g., innovative materials design, it is often necessary to obtain a description of the interface at the molecular scale. 
For example, the selectivity and efficiency of advanced membrane materials is determined by factors that are directly linked to the molecular structure of the interface between the stationary and mobile phases. \cite{konatham_simulation_2013,cohen-tanugi_water_2012} Similarly, the operation of heterogeneous catalysts is determined by the environment of the active site at the atomistic level. \cite{hutchings_spiers_2021}
Even in the ideal case of a perfectly planar surface in contact with a liquid, intermolecular interactions lead to highly non-ideal regions of the fluid phase. \cite{finney_electrochemistry_2021}
Measuring the properties and structure of this interfacial region is challenging and often requires specialist experimental techniques \cite{garcia_dynamic_2002,hayes_double_2011,vlieg_surface_1988,collins_ellipsometry_1990,watts_optical_2019}
This is compounded in the case of changing or fluctuating surfaces.
Crystal growth is a paradigmatic example of such a macroscopic process emerging from the collective behavior of a large number of molecules and the evolution of a solid/liquid interface.
Crucially, crystallization is affected by mechanistic, thermodynamic and kinetic details regarding the fundamental physicochemical steps occurring at the scale of single growth units. \cite{mullin_crystallization_2001,anwar2011uncovering,salvalaglio2012uncovering,han2019solvent} 

Given the challenges alluded to above, it is commonplace in many fields to employ molecular simulation techniques, often based on Molecular Dynamics (MD), to gain atomistic-level insight into the structure and dynamics of collective phenomena at the solid/liquid interface.\cite{frenkel2001understanding,de2002molecular,allen_computer_2017,palmer2015recent} 
MD simulations provide direct access to the length- and time-scales for molecular processes which are often difficult to probe in experiments. In particular, MD simulations are ideally suited to describe equilibrium conditions in dense systems such as single and multi-component liquids.\cite{allen_computer_2017} Technological applications, however, often exploit out-of-equilibrium, concentration-driven processes to achieve their function. 
Crystal precipitation represents once again a particularly relevant example in this context. Crystallization processes in fact operate out-of-equilibrium, and are governed by the kinetics of fundamental steps such as nucleation and growth.\cite{kashchiev_nucleation_2000,mullin_crystallization_2001,piana2005simulating}
Ideally, MD simulations aimed at understanding the control of solid/liquid interfaces on these steps need to be able to capture the evolution of an out-of-equilibrium system under well-defined conditions of temperature, pressure and supersaturation. 
When dealing with out-of-equilibrium, concentration-driven molecular simulations of systems containing more than one phase, however, the accessible system sizes associated with MD (due to finite computational resources) can limit the scope of such techniques to tackle the problems of interest, and often requires the development of dedicated theoretical descriptions and ad-hoc corrections.\cite{reguera2003phase,wedekind2006finite,grossier2009reaching,salvalaglio2013controlling,agarwal2014solute,salvalaglio2015molecular,salvalaglio2016overcoming}

The solid/liquid interface in macroscopic systems is in contact with a bulk liquid phase that changes its properties on timescales that are significantly longer than those typically associated with molecular-scale phenomena.
For example, in the case of multi-component solutions, bulk solute concentrations can be considered unchanging on the timescales required for solute monomer incorporation into a crystal growth site at the surface. A faithful representation of these types of processes in MD requires an ability to model systems out-of-equilibrium as a function of constant bulk solution properties. 
Achieving this is far from trivial, because it usually requires simulations with open boundaries to allow exchange of species at the solid/liquid interface with the `surrounding bulk' to maintain steady-state out-of-equilibrium conditions.
Suitable techniques to perform these kinds of simulations include hybrid grand-canonical Monte Carlo/MD,
in which there is a dynamic exchange of molecular species between the simulation cell and an external, virtual reservoir defined by a constant chemical potential.\cite{ccaugin1991molecular,papadopoulou1993molecular,yau1994contact,lo1995alternative,lynch1997grand,delgado2003usher}
Despite their elegant formalism, hybrid MC/MD approaches are extremely inefficient when dealing with dense fluids, as the probability of acceptance of exchange moves between a molecule in the fluid and the external reservoir becomes very small. 
An alternative approach, known as Constant Chemical Potential Molecular Dynamics (C$\mu$MD), was recently introduced to overcome these limitations that mimics open-boundary, constant composition, out-of-equilibrium conditions.\cite{perego2015molecular,ozcan2017concentration,radu2017enhanced,karmakar2018cannibalistic,loganathan2019understanding,namsani2019direct,ozcan2020modeling} 
C$\mu$MD, similarly to adaptive resolution simulation methods\cite{fritsch2012adaptive,potestio2013hamiltonian,wang2013grand,agarwal2015molecular}, introduces an internal reservoir of molecules in the fluid phase, and applies ad-hoc forces to regulate the flux to/from the reservoir and the `bulk' solution in contact with the interface.
This technique has already proven to be very effective to understand the molecular-scale processes associated with crystallization, surface adsorption and permeation, and its application is not limited by the density of the fluid phase.\cite{perego2015molecular,han2019solvent,ozcan2020modeling,finney_electrochemistry_2021}

Adopting C$\mu$MD to gain atomistic-level insight into the solid/liquid interface allows for a direct comparison of the structure and dynamics of interfacial species with the predictions of mean field models (based on constitutive equations) to describe mass transfer at mesoscopic and macroscopic length scales.
In order to achieve this goal, a fully consistent analysis and interpretation of atomistic-level information in the context of constitutive mass-transfer equations is necessary. 

The remainder of this paper is organised as follows: firstly, using conceptual examples we illustrate how accounting for microscopic interactions between the surface and the fluid phase is essential to capture the complexity of concentration fields at the nanometre scale with constitutive equations; secondly, we employ constitutive equations to interpret the concentration profiles obtained studying the adsorption of NaCl(aq) at graphite and NaCl(s), and extract the ion-specific excess free energy of interaction with the solid surface. The non-ideality of the solution structure stemming from asymmetric adsorption of species with different charge is responsible for local violations of electroneutrality in solution, and is key to develop consistent mesoscopic models of the solid/liquid interface.

\begin{figure}[th]
    \centering
    \includegraphics[width=0.45\textwidth]{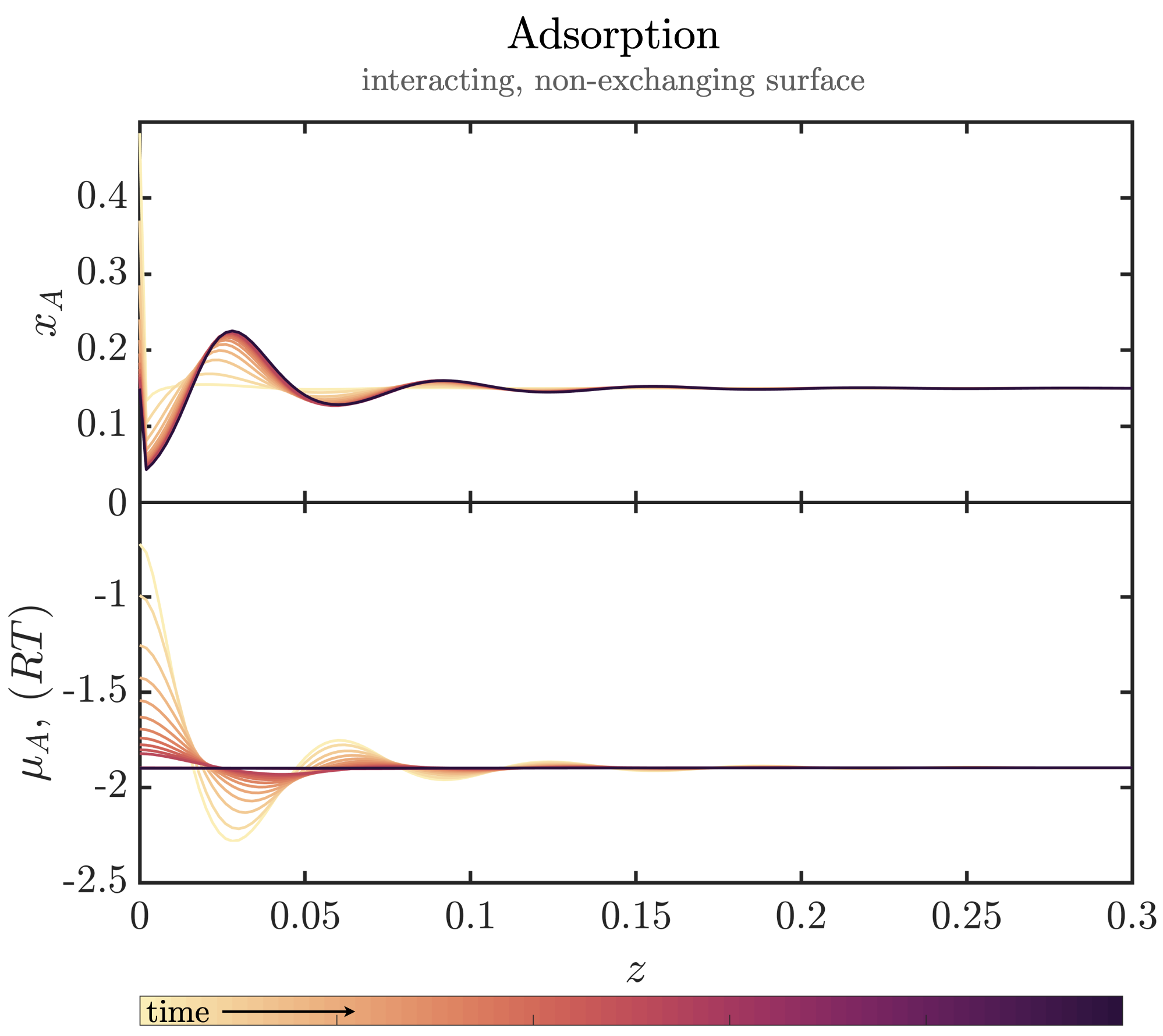}
    \caption{Conceptual model of mass transfer in close proximity to an interacting, non-exchanging surface. Time-dependent solution of the microscopic material balance for a binary liquid mixture in contact with an interacting surface in a mono-dimensional domain. \emph{Left.} Molar fraction profile of the solute (A) as a function of time. At a steady state, the solute accumulates at the interface according to the potential of mean force for adsorption at the interface. \emph{Right.} The driving force $\nabla{\mu_A}$ becomes null at a steady state, where the $\mu_A$ profile is flat and the flux is null. These profiles were obtained by solving the time-dependent diffusion equation, with initial conditions $x_A(z,0)=0.15$, and boundary conditions $x_A(0.5,t)=0.15$ and $\frac{\partial{\mu_A}}{\partial{z}}\vert_{z=0,t}=0$.}
    \label{fig:adsorption1}
\end{figure}

\subsection{Microscopic inconsistencies in models of transport at interfaces: Film Theory}
To highlight the need for a reconciliation between meso- and molecular-scale representations of transport processes at interfaces, we start by analysing Film Theory (FT). FT is a typical simplification of the mass transfer constitutive equations applied to solid/liquid interfaces that leads to an inconsistent description of the composition field in the liquid phase in the presence of strong interactions between the solid and fluid phases.
The mass transfer flux at the solid/liquid interface in FT is expressed as\cite{mullin_crystallization_2001}
\begin{equation}
J_i=-k\left(c_i^s-c_i^b\right)
\end{equation}
where $J_i$ is the molar flux of species $i$, $k$ is a mass transfer coefficient, and $c_i^s$ and $c_i^b$ are the concentrations of species $i$ at the solid surface and in the bulk of the fluid phase, respectively. The above equation is predicated on two key assumptions: i) the concentration gradient is linear within a fluid film of thickness $\delta$ in contact with the surface; and, ii) the driving force for the diffusive mass transfer of $i$ is a gradient in concentration.
FT is grounded in Fick's diffusion equation, and the mass transfer coefficient is defined as the ratio between the Fick diffusion coefficient, $\mathscr{D}_i$, and $\delta$, which can be obtained from adimensional number correlations.\cite{mullin_crystallization_2001}
FT provides a simple framework to express diffusion dominated flowrates across interfaces as a function of the concentrations described above where $c_i^s$ is typically considered equal to the equilibrium concentration.
While this model is able to capture the qualitative behavior of solute fluxes in crystal growth processes, it demonstrates fundamental inconsistencies when compared with results from simulations and experiments able to resolve molecular-level length and time scales. 
Attractive and repulsive intermolecular interactions between the surface and solute species can lead to concentrations at the interface that deviate from $c_i^eq$, thus introducing inconsistencies between the sign of the term $\left(c_i^s-c_i^b\right)$ and the direction of the net diffusive flux at the interface. Indeed, the concentration profile postulated by FT is verified only in the case of a weakly interacting surface, that does not induce significant solute adsorption at the solid/liquid interface.

\section{A consistent description across scales: Maxwell-Stefan diffusion.}
\noindent A fully consistent description of diffusive mass transfer at attractive/repulsive interfaces requires a framework in which the chemical potential gradient is explicitly considered as the driving force for the flux of species. Such a framework is provided by the generalised Maxwell-Stefan (MS) diffusion equation. In this section, we progressively build the necessary complexity within the MS diffusion equation to consistently interpret the effect of a strongly interacting surface on the multi-component, electrolyte solution modelled via fully-atomistic C$\mu$MD simulations. 

\subsection{Interacting, non-exchanging surface.}
\noindent We begin by considering the adsorption of an electrically neutral solute A in solvent B interacting with a solid surface unable to exchange mass with the fluid phase. The MS equation for the diffusive flux of A in this system is\cite{prausnitz1998molecular,liu2011predictive,liu2013diffusion}
\begin{equation}
J^*_A=-c_A\frac{\dH_{A,B}}{RT}\nabla{\mu_A}
\label{eq:MS}
\end{equation}
where $J_A^*=x_A(v_A-v^*)$, with $v_A$ and $v^*$ being the velocity of component $A$ and the molar weighted average velocity, respectively.
$c_A$ in the above equation is the molar concentration of $A$, while $\dH_{A,B}$ is the MS diffusion coefficient, and $\nabla{\mu_A}$ is the chemical potential gradient of A. 

The species-specific intermolecular forces established between the surface and each component of the fluid phase can be introduced into the MS diffusion equations through $\nabla{\mu_A}$ by explicitly considering the dependence of $\mu_A$ on the distance from an interacting surface. 
The chemical potential of $A$ can be written as a function of its molar fraction, $x_A(\mathbf{r})$, and its activity coefficient, $\gamma_A(\mathbf{r})$ as: 
\begin{equation}
\mu_A(\mathbf{r})=\mu_A^0 + RT\ln{x_A(\mathbf{r})}+RT\ln{\gamma_A(\mathbf{r})}
\label{eq:muA}
\end{equation}
where $\mu_A^0$ is a reference chemical potential under standard conditions, and $RT\ln{x_A(\mathbf{r})}$ is the free energy of an ideal mixture of composition $x_A$ ($R$ and $T$ being the ideal gas constant and temperature, respectively). The term $RT\ln{\gamma_A(\mathbf{r})}$ defines $g_A^E(\mathbf{r})$, i.e. the partial molar excess free energy of A in solvent B, at distance $\mathbf{r}$ from the interacting surface\cite{prausnitz1998molecular}. 
In contrast with the excess free energy of component A in the bulk of a real fluid mixture, which is a function of the composition only (in the absence of an external field), the excess free energy term in the presence of an interface accounts for the asymmetric intermolecular forces between the species in solution and the solid surface, and thus depends on the distance from the interface. Equation \ref{eq:MS} can be rewritten as: 
\begin{equation}
J^*_A=-c\dH_{A,B}\left[\nabla{x_A}+\frac{x_A}{RT}\nabla{g_A^E(\mathbf{r})}\right]
\label{eq:MS_wall}
\end{equation}
$\nabla{g_A^E(\mathbf{r})}$ can be interpreted as the mean thermodynamic force exerted by the solid surface on species $A$, which is responsible for the emergence of features in the local composition field in the vicinity of a solid surface. 
In the case of an inert surface, the excess chemical potential of A in the fluid becomes independent from $\mathbf{r}$, i.e. $\nabla{g_A^E(\mathbf{r})}=0$ and there is no accumulation/depletion of solute in the fluid phase near the interface. In such conditions, Fick diffusivity is recovered as $\mathscr{D}_{A,B}=\dH_{A,B}\left(1+\frac{\partial{\ln{\gamma_A}}}{\partial\ln{x_A}}\right)$, and simplified approaches like FT will provide yield a coherent representation of the concentration field. Furthermore, in the case of an ideal mixture (i.e. $\gamma_A=1$), ${\mathscr{D}_{A,B}=\dH_{A,B}}$. 

When the concentration field in the presence of a heterogeneous interacting surface is stationary and $J^*_A=0$, the excess thermodynamic force $\nabla{g_A^E(\mathbf{r})}$ can be computed from the expression:  
\begin{equation}
\nabla{g_A^E(\mathbf{r})}=-\frac{RT}{x^{ss}_A(\mathbf{r})}\nabla{x_A^{ss}(\mathbf{r})}
\label{eq:excessFE}
\end{equation}
where $x_A^{ss}$ is the steady state molar fraction field. Equation \ref{eq:excessFE} is equivalent to $g_A^E(\mathbf{r})=-{RT}{\ln{x_A^{ss}}}+C$, where $C$ is an arbitrary integration constant. The excess free energy term $g_A^E(\mathbf{r})$ can thus be computed directly from molecular simulations by sampling the steady-state composition field of species A as a function of distance from the solid/liquid interface.  
As shown in Fig.\ref{fig:adsorption1}, solving the microscopic material balance accounting for $\nabla{g_A^E(\mathbf{r})}\neq{0}$ in a mono-dimensional domain leads to a microscopically consistent concentration field. In the case of an attractive surface, from an initial homogeneous concentration of $A$, accumulation of solute at the surface occurs at equilibrium. The steady state solution of this mass balance provides the correct equilibrium conditions, i.e.  $\nabla{\mu_A}=0$, while $\nabla{x_A^{ss}(\mathbf{r})}\neq{0}$. 

\subsection{Interacting, exchanging surface}
Let us now consider the example of a solid surface exchanging molecules of solute (A) with a two-component solution constituted by the electrically neutral solute A and solvent B. This example applies to crystal/solution interfaces out-of-equilibrium in which solute molecules (A, in this example) are either released by, or incorporated into the solid surface. 

\begin{figure*}[th]
    \centering
    \includegraphics[width=0.9\textwidth]{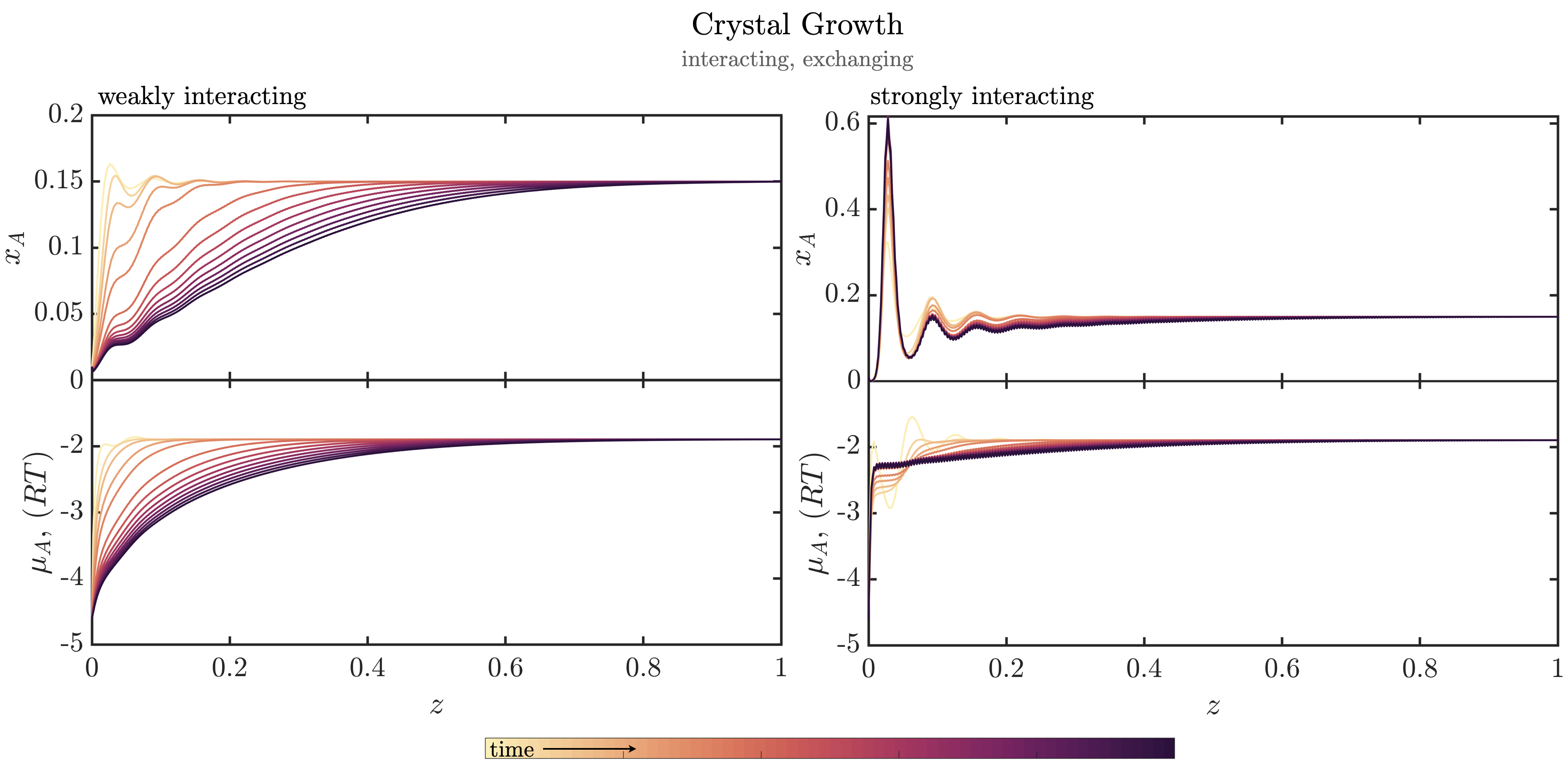}
    \caption{Conceptual model of mass transfer in the proximity of an out-of-equilibrium, interacting and exchanging surface provided by the time-dependent solution of the microscopic material balance for a binary liquid mixture in a mono-dimensional domain. \emph{Top: weakly interacting surface} Molar fraction and chemical potential profiles of the solute (A) as a function of time. In this case the gradient of concentration and chemical potential have the same sign, and the use of the simplified description of mass transfer from Film Theory (FT) provides a reasonable description of the microscopic concentration profile. 
    \emph{Bottom: strongly interacting surface} In the case of a strongly interacting surface, the microscopic description of concentration and chemical potential in the interfacial region is inconsistent with the descriptions from FT. For example, the concentration difference between interface and bulk has the opposite sign of the chemical potential difference, and simplified models cannot capture a microscopically consistent composition field.}
    \label{fig:growth1}
\end{figure*}
In the current case, the diffusive flux of A in proximity to a crystal surface is described by the MS diffusion equation reported in Equation \ref{eq:MS_wall} discussed in the context of an interacting surface. However, in order to capture the flux of A across the solid/liquid phase boundary, the microscopic mass balance should reflect the fact that, at constant temperature and pressure, the chemical potential of a crystal comprised of A monomers, $\mu_A^{xtal}(T,P)$, is constant.
One can therefore express the boundary conditions in terms of chemical potentials of the solid and liquid species. 
Considering that the diffusion driving forces include the contribution of $g_A^E(\mathbf{r})$, as introduced in Equation \ref{eq:muA}, the microscopic material balance reads: 
\begin{equation}
\frac{\partial{x_A}}{\partial{t}}=\frac{1}{RT}\nabla{\left[x_A\dH_{A,B}\nabla\left(RT\ln{x_A}+g_A^E(\mathbf{r})\right)\right]}
\end{equation}
with the boundary condition at the solid/liquid interface: 
\begin{equation}
\mu\vert_{z_{int}}=\mu^{xtal}(T,P)=RT\ln(x_A^*\gamma_A^*)
\end{equation}
where $x_A^*$ and $\gamma_A^*$ are the molar fraction and activity coefficient of component A at equilibrium in the bulk solution. 
As shown in Fig. \ref{fig:growth1}, for a weakly interacting surface where the contribution of $\nabla{g_A^E(\mathbf{r})}$ to the mass transfer driving force is small compared to the contribution of $\nabla{x_A^{ss}(\mathbf{r})}$, the composition field remains qualitatively consistent with the predictions of simple models for mass transfer at interfaces. However, when the contribution of $\nabla{g_A^E(\mathbf{r})}$ becomes significant (see Fig. \ref{fig:growth1}, strongly interacting surface) the composition profile shows the emergence of a local solution structure with solute accumulation at the interface, incompatible with models such as FT. 

\section{Excess surface free energy from C$\mu$MD simulations} 
Having introduced a framework to interpret complex composition profiles, in this section we extend our description to analyse results from C$\mu$MD simulations of interacting surfaces in contact with NaCl(aq) electrolyte solutions with varying solute concentrations. We consider first the case of a non-exchanging surface (graphite), and then a surface exchanging ions with the solution (\{100\} NaCl(s)).

\begin{figure*}[th]
    \centering
    \includegraphics[width=\textwidth]{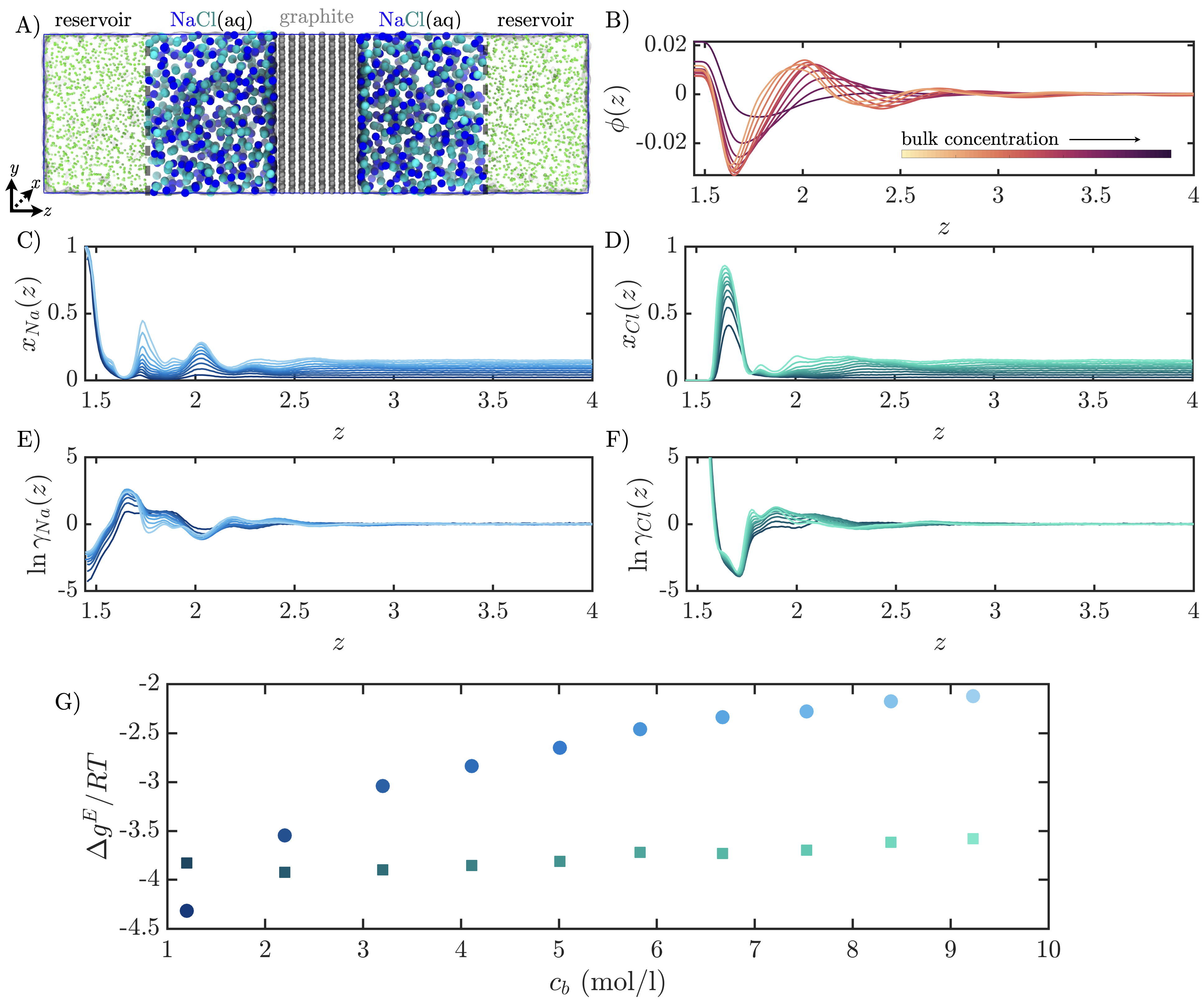}
    \caption{C$\mu$MD Simulations of graphite interfaces in contact with NaCl aqueous solutions highlighting the asymmetric adsorption of ions ($\sim$1-10M). A) A representation of the simulation box setup, displaying the ion reservoir in green, and the position at which C$\mu$MD forces are applied as a dashed line. B) Electric potential $\Phi(z)$ generated by the local ion charge fluctuations in solution. C-D) Na$^+$ and Cl$^-$ molar fraction profiles. E-F) Excess chemical potential associated with the asymmetric interaction of  Na$^+$ (E) and Cl$^-$ (F) ions with the graphite surface.G) Excess adsorption free energy for Na$^+$ (circles), and Cl$^-$ (squares) as a function of the concentration in the solution bulk $c_b$. The color code corresponds to $c_b$ and is consistent with those in panels C-F.  }
    \label{fig:graphite}
\end{figure*}

\subsection{NaCl(aq)/graphite interfaces.}
\label{sec:equilibrium}
The concentration field obtained in atomistic simulations of a graphite-electrolyte solution interface (see Fig.\ref{fig:graphite}~A) is a result of the combined effect of differing affinities for ionic species to accumulate at the interface, and of the emergent electric fields in this region generated by local violations of electroneutrality associated with the asymmetric adsorption. \cite{finney_electrochemistry_2021} The system considered in this section is represented in Fig.\ref{fig:graphite}~A. As such, additional driving forces, such as an electric field (which is equal to the gradient of the electric potential field, $\nabla{\phi}$), can be introduced into the MS diffusion equation to describe the global diffusive flux of cations and anions.\cite{bird2002transport}
The MS equations for Na$^+$ and Cl$^-$ in solution, respectively read: 
\begin{dmath}
\frac{N_{Na}x_{Cl}-N_{Cl}x_{Na}}{c\dH_{NaCl}}+\frac{N_{Na}x_{W}-N_{W}x_{Na}}{c\dH_{NaW}}=
-\nabla{x_{Na}}-x_{Na}\frac{\nabla{g^E_{Na}}}{RT} - \frac{\rho_{Na}z_{Na}F}{cRT\,M_{Na}}\nabla{\phi}
\label{eq:MSNa}
\end{dmath}
and:
\begin{dmath}
\frac{N_{Cl}x_{Na}-N_{Na}x_{Cl}}{c\dH_{NaCl}}+\frac{N_{Cl}x_{W}-N_{W}x_{Cl}}{c\dH_{ClW}}=
-\nabla{x_{Cl}}-x_{Cl}\frac{\nabla{g^E_{Cl}}}{RT} - \frac{\rho_{Cl}z_{Cl}F}{cRT\,M_{Cl}}\nabla{\phi}
\label{eq:MSCl}
\end{dmath}
where $N_i$, $z_i$, $\rho_i$ and $M_i$ indicate the molar flux, density, valency and molar mass of species $i$, respectively.\cite{bird2002transport} 
It should be noted that in the case of an uncharged surface, the terms dependent upon $\nabla{\phi}$ on the right-hand side of Equations \ref{eq:MSNa} and \ref{eq:MSCl} emerge solely due to a non-zero net charge distribution in the fluid phase close to the surface; indeed $\phi$ is computed as: 
\begin{dmath}
\phi(z)=\int_0^{z^\prime}-\frac{Fc}{\epsilon_0\epsilon_r}\left[\int_0^{z^\prime}x_{Na}dz-\int_0^{z^\prime}x_{Cl}dz\right]dz
\label{eq:psi}
\end{dmath}
where $F$ is Faraday's constant, $c$ is the bulk concentrations of ions, and $\epsilon_0$ and $\epsilon_r$ are constants for the vacuum and relative permittivity of the solution medium.

The source of the asymmetric adsorption of ions of opposite electrical charge is the different potentials of mean force that characterise the surface-solution interactions, and which are captured by the excess free energy term $g^E_i(\mathbf{r})$ for ion $i$. 
We can exploit the steady state conditions obtained from open boundary MD simulations to extract the mean force terms $\mathscr{W}_i(\mathbf{r})$, which balances the combined effect of the electric field and that of the chemical potential of the ideal solution, leading to a constant electrochemical potential at equilibrium: 
\begin{dmath}
\nabla{g_i^E}=-\frac{RT}{x_i}\nabla{x_i} - \frac{\rho_{i}z_{i}F}{c_i\,M_{i}}\nabla{\phi}
\label{eq:meanforce}
\end{dmath}
As shown in Fig. \ref{fig:graphite}~C and D, the molar fraction profile of both cation and anion can be extracted from simulations, and the electrostatic potential computed using Eq. \ref{eq:psi}, as shown in Fig. \ref{fig:graphite}~B. The profile of the ion-specific excess free energy of adsorption $g_i^E/RT=\ln{\gamma_i(\mathbf{r})}$, reported in Fig. \ref{fig:graphite}~E and F is thus computed using Eq. \ref{eq:meanforce}. The $g_i^E/RT$ profiles overall capture the asymmetric driving forces for ion adsorption at graphite surfaces responsible for non-ideal character of the electrical double layer in these systems\cite{finney_electrochemistry_2021}. The excess free energy of adsorption of the cation has a minimum in direct contact with the graphite surface, while the anion excess free energy has a minimum that corresponds to an adjacent layer. 

We define the excess free energy of adsorption as, $\Delta{g^E_i}=\min{g^E_i(z)}-g^E_i(z_b)$ for ionic species $i$, where $z_b$ is a position in the bulk.
$\Delta{g^E_{Na}}$ and $\Delta{g^E_{Cl}}$ exhibit a different dependence on the value of bulk concentration as shown in Fig. \ref{fig:graphite}~G. While $\Delta{g^E_{Na}}$ becomes less negative with increasing bulk concentrations, $\Delta{g^E_{Cl}}$ remains fairly independent from bulk concentration. 
The free energy change for the adsorption of single cations at the carbon surface is greater than for chloride ions; \cite{williams_effective_2017} hence, the result shown in Fig. \ref{fig:graphite}~G for the lowest concentration.
At high bulk concentrations, however, crowding of solute species at the interface leads to significant ordering of the ions, particularly so for cations in the first solution layer. 
The relative accumulation of ions in this region is  reduced relative to the bulk levels as concentration increases, and minima for the adsorption of cations in the second and third solution layers beyond the surface (see Fig. \ref{fig:graphite}~E) deepen.

\begin{figure*}[th]
    \centering
    \includegraphics[width=\textwidth]{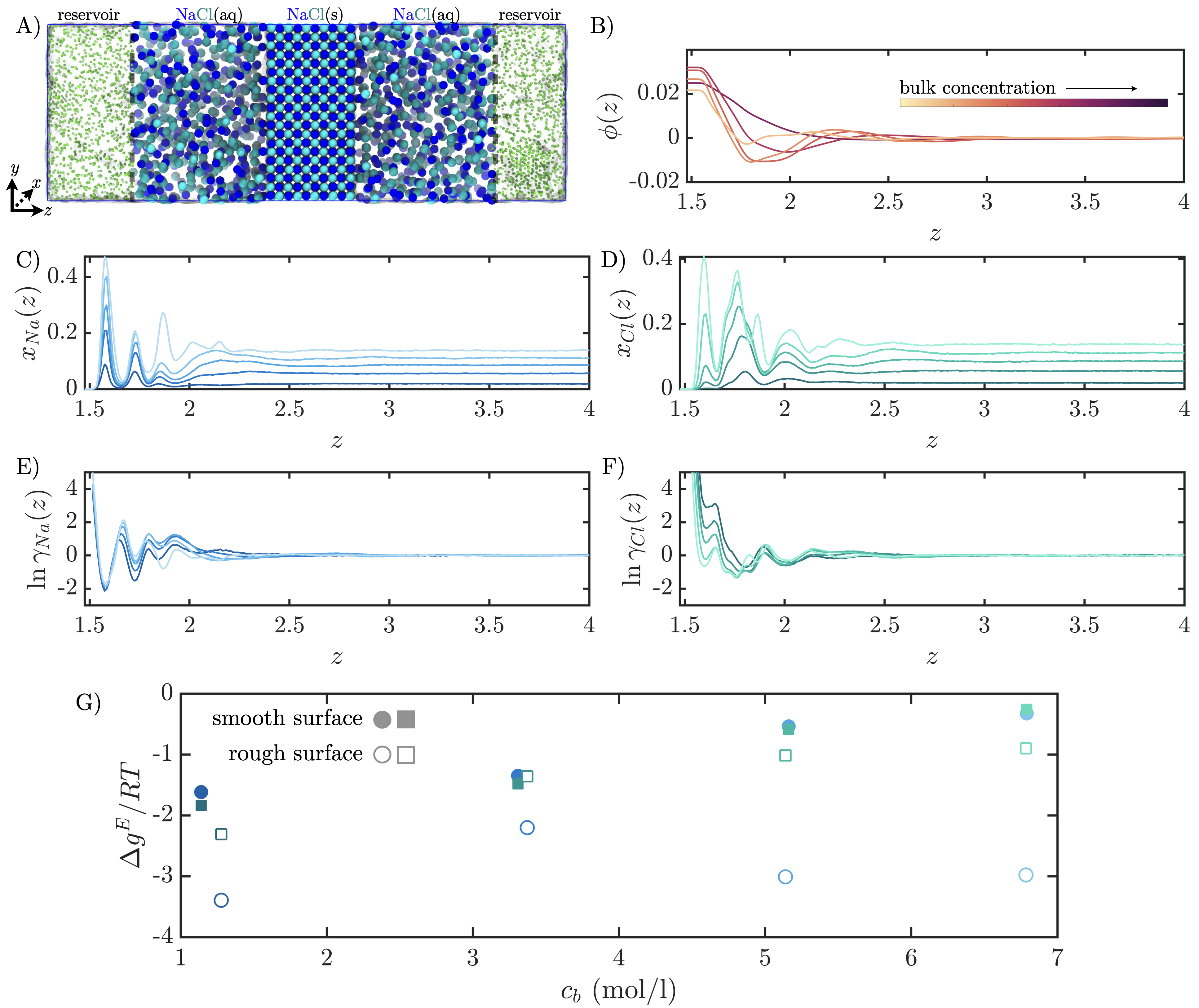}
    \caption{C$\mu$MD Simulations of NaCl(s) \{100\} surfaces in contact with NaCl aqueous solutions (~1-7M). A) A representation of the simulation box setup, displaying the ion reservoir in green, and the position at which C$\mu$MD forces are applied as a dashed line. B) Electric potential $\Phi(z)$ generated by the local charge fluctuations due to the asymmetric adsorption of ions at the interface. C-D) Na$^+$ and Cl$^-$ molar fractions profiles, displaying an asymmetric behaviour in proximity to the system surface. E-F) Excess chemical potential associated with the asymmetric interaction of  Na$^+$ (E) and Cl$^-$ (F) ions with the graphite surface.G) Excess adsorption free energy for Na$^+$ (circles), and Cl$^-$ (squares) as a function of the concentration in the solution bulk $c_b$. The color code corresponds to $c_b$ and is consistent with those in panels C-F. Full symbols refer to smooth NaCl(s), while empty symbols correspond to values of the adsorption excess free energy for rough surfaces.}
    \label{fig:growth2}
\end{figure*}
\subsection{NaCl(aq)/NaCl(s) interfaces.}
\label{sec:non-equilibrium}
Finally, we analyse out-of-equilibrium NaCl(s) \{100\} crystalline interfaces in contact with a bulk solution at constant composition.
When the system evolves under diffusion-limited growth, the total flux of ions to and from the crystal is determined by the gradient in chemical potential. In this case, ions at the interface are not restricted to the liquid phase and can be incorporated into the evolving crystalline surface. Their chemical potential displays a discontinuity at the boundary between the solid and liquid phases.  
Together with the chemical potential, the system's composition, density and local structure also display discontinuities at the solid/liquid boundary. 

When the solution bulk is supersaturated, i.e., $\mu_{NaCl}^{aq}>\mu_{NaCl}^{xtal}$ (where superscript $aq$ and $xtl$ indicate the aqueaous and crystal chemical potential, respectively), the solid/liquid interface will act as a sink for the removal of ions from the liquid phase at a rate determined by the diffusion of ions. Conversely, in understaurated conditions, where $\mu_{NaCl}^{aq}<\mu_{NaCl}^{xtal}$, the solid/liquid interface supplies ions to the bulk solution. 
In order to investigate the role of surface structure on the ion-specific excess free energy, we carry out simulations on both perfectly planar (smooth) and rough surfaces where at least three crystal terraces are exposed to solution initially (see Methods).    
As in the case of graphite, C$\mu$MD simulations reveal a solution structure heavily affected by specific surface-solute interactions. 
For both smooth and rough surfaces, the solution composition profiles display a layered solute structure at the interface, similar to the one observed in Fig. \ref{fig:graphite}. Also in this case, the cations in solution tend to accumulate in the first adsorbed solute layer in contact with the surface (see Fig. \ref{fig:growth2}~C and D), leading to local violations of the solution electroneutrality perpendicular to the surface and a gradient in the electric potential (Fig. \ref{fig:growth2}~B). 

At smooth NaCl(s) surfaces the cation excess is readily compensated by an anion excess in the immediately adjacent solution layer. In this limit, even if the accumulation of ions from the solution is asymmetric, the strength of the adsorption interaction of both ions with the surface is comparable. This is quantitatively reflected in the adsorption excess free energy (Fig. \ref{fig:growth2}~E-G) that is essentially the same for both ions and exhibits the same dependence on the bulk solute concentration, with increasing bulk ion concentrations leading to a less negative $\Delta g^E$. 
A more complex behaviour emerges in the case of roughened \{100\} NaCl crystal surfaces. At the rough surface the outermost crystal layer can exchange ions with the liquid phase much more readily on the timescale of the simulations than occurs at perfectly planar crystal surfaces of NaCl. 
Moreover, under-coordinated surface ions tend to develop stronger interactions with molecules in the fluid phase than ions in the `bulk' of the crystal. 
The combined effect is a significant reorganisation of both the liquid and the solid at the interface.
The additional surface area and exposed step edges lead to increased ordering of ions in first few solution layers, cf. the smooth NaCl surface, with fluctuations to the screening of the charge associated with a preferentially adsorbed first sodium layer.
This observation is consistent with the excess adsorption free energy trend as a function of bulk concentration on rough surfaces (see Fig. \ref{fig:growth2}~B). 
While the excess adsorption free energy of the anion on rough surfaces is only marginally lower than that of both the anion and cation on smooth surfaces, $\Delta{g^E_{Na}}$ is significantly lower, and does not show a clear dependence on the bulk ionic concentration. 
This result suggests that interactions with under-coordinated sites at the interface is an important contributing factor towards ion accumulation.

\subsection{Outlook: linking dynamics across the scales} 
\begin{figure*}[th]
    \centering
    \includegraphics[width=\textwidth]{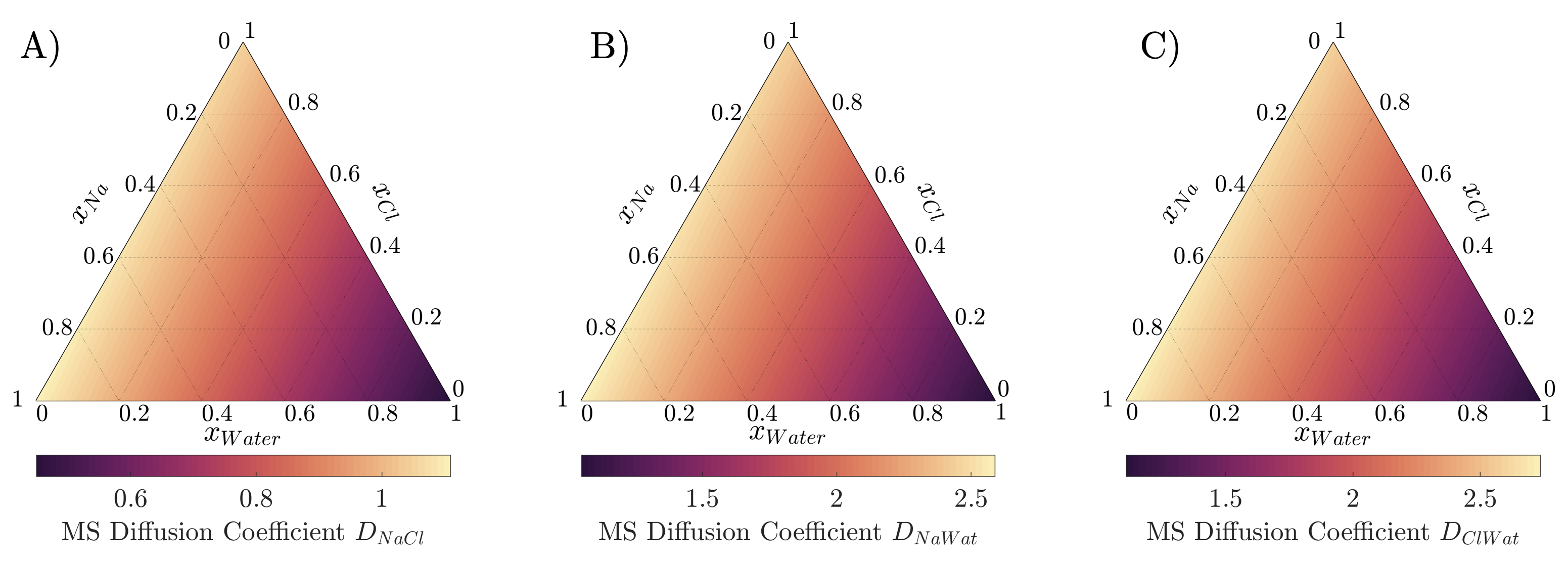}
    \caption{Triangle plots representing variations in the MS diffusion coefficients computed using the Darken equation (Eq. \ref{eq:darken}) for NaCl(aq). Since close to the S/L interface electroneutrality is locally violated, and molar fractions of sodium and chloride ions should be considered independent variables.}
    \label{fig:DiffusionCoeff}
\end{figure*}

In the above discussion we have shown how Maxwell-Stefan diffusion theory can be applied to extract quantitative information from both equilibrium and out-of-equilibrium molecular simulations of interfaces subject to a constant background concentration, at steady state. $g_i^E(\mathbf{r})$ - the excess free energy of interaction between species $i$ and an interface located at distance $\mathbf{r}$ - is a key component responsible for the emergence of local deviations from ideal solution behavior. 
With knowledge of $g_i^E(\mathbf{r})$, one can inform the development of fully consistent mean field models based on MS theory to capture also the evolution of the liquid phase composition close to an interface. The information needed to achieve this goal is the set of $N(N-1)/2$ MS diffusion coefficients, where $N$ is the number of species in the system. As discussed in detail in the comprehensive review of Liu et al. \cite{liu_diffusion_2013} the MS diffusion coefficients $\dH_{i,j}$ can be computed using a variety of different simulation approaches. In particular we highlight that the application of the multi-component Darken equation \cite{darken_diffusion_1948,liu2011predictive,liu2011fick,liu2012fick} allows for a straightforward calculation of $\dH_{i,j}$ from the self diffusion coefficient $D_{i,self}$ and from the component molar fraction $x_i$: 
\begin{equation}
\dH_{i,j}=D_{i,self}D_{j,self}\sum_{i=1}^N\frac{x_i}{D_{i,self}}
\label{eq:darken}
\end{equation}
In the context of this study we have performed dedicated bulk simulations to obtain the self-diffusion coefficients of Na$^+$, Cl$^-$, and SPC/E water (see the Methods section). Nevertheless, we highlight that calculation of the self-diffusion coefficients can be performed on the same trajectory obtained with C$\mu$MD, as done in Ref. \cite{finney_electrochemistry_2021} to assess the ions' mobility as a function of distance from the solid/liquid interface. 
Since in this work we are considering mass transfer in the vicinity of interacting surfaces that exert an asymmetric effect on different species, thus leading to local violations of typical constraints such as electroneutrality (Fig. \ref{fig:graphite} and \ref{fig:growth2}B), in the application of the multicomponent Darken equation we need to explicitly account for three independent molar fractions. The MS diffusivities $\dH_{i,j}$ computed as function of three independent molar fractions subject to the stoichiometric constraint, are reported in the triangular plots represented in Fig. \ref{fig:DiffusionCoeff}. In particular, we note how local deviations from the ideal solution behavior observed close to the surface may lead to fluctuations by up to a factor two in the MS diffusivities. 
Depending on the level of detail required for the solution of a mean field model informed by MD simulations, this aspect may require consideration as it will likely impact on the computational complexity of the model. 

\section{Conclusions}
In this work we demonstrated how molecular-level information obtained from C$\mu$MD\cite{perego_molecular_2015} simulations of solid-liquid interfaces can be interpreted consistently and quantitatively using the Maxwell-Stefan theory of diffusive mass transfer. 
This enables a characterisation of non-idealities in two fundamental processes at the heart of a wide range of chemical technologies, namely:
\begin{itemize}
    \item The equilibration of a fluid phase in contact with a non-exchanging, interacting surface (Sec.\ref{sec:equilibrium}).
    \item The steady-state evolution of an out-of-eqilibrium, interacting surface able to exchange molecules with a fluid phase under the effect of a constant driving force (Sec. \ref{sec:non-equilibrium})
\end{itemize}
The former process drives adsorption-based technologies; the latter is fundamental to understand and control crystallization.
The fully constitutive approach is based on the calculation of an excess free energy profile as a function of distance from a surface, which is otherwise extremely challenging to compute using alternative simulation approaches.
This allows us to demonstrate how, in the presence of strong interactions between solid- and fluid-phase species, deviations in the spatially-varying chemical potential and concentration fields differ from the predictions of simplified models of mass transport at interfaces.
A consistent mean-field model, like the one presented here, is key to efficiently transfer information across the scales and investigate out-of-equilibrium, concentration-driven processes such as crystallization in complex surface geomteries at the nanometer scale over a wide range of bulk solute concentrations.

Our analysis, applied to an extensive set of NaCl(aq) simulations in contact with graphite (Sec. \ref{sec:equilibrium}) and NaCl crystal surfaces (Sec. \ref{sec:non-equilibrium}), reveals a ubiquitous asymmetric adsorption behavior. In particular, we consistently observe the preferential adsorption of the cation, driven by a more favourable excess adsorption free energy both on graphite as well as on NaCl(s) surfaces. The asymmetric adsorption behaviour leads to local violations of the solution electroneutrality and to the emergence of an electric potential in the fluid phase in contact with the crystal surface analogous to the double layer potential that NaCl(aq) develops in contact with graphite electrodes.

\section{Methods}
\subsection{C$\mu$MD simulations.}
\noindent The simulations herein were performed using the GROMACS 2018.6 package \cite{hess_gromacs_2008} and the PLUMED plugin (version 2.5) \cite{tribello_plumed_2014}.
Unless otherwise stated, C$\mu$MD simulations \cite{perego_molecular_2015} were carried out using a leapfrog time integration algorithm with a 2~fs timestep. 
Simulations were performed in the \textit{NVT} ensemble at 298~K; the temperature was held constant within statistical fluctuations using the Bussi-Donadio-Parrinello thermostat. \cite{bussi_canonical_2007}
Periodic boundaries were imposed in three-dimensions to the orthorhomibic simulation cells shown in e.g., FIG. \ref{fig:growth1}~A.
Electrostatic interactions were treated using Particle Mesh Ewald  summation\cite{darden_particle_1993} with a cut-off applied to the calculation of the real-space contribution from atoms within 0.9~nm.
A truncation was also applied to the calculation of pairwise Van der Waal interactions which was limited to those within 0.9~nm, and with a potential shift applied to minimise the fluctuations in the forces around the cut-off distance.
The LINCS algorithm \cite{hess_lincs:_1997} was adopted to constrain the structure of water molecules.

In the C$\mu$MD algorithm adopted in this work, a continuous force, $F^{\mu}$, is applied to ions in the vicinity of a fixed distance from a reference point in the simulation cell, so as to maintain target ion number densities in control regions (CRs), $n^{\mathrm{CR}}$, beyond the solid/liquid interface:
\begin{equation}
    F_i^{\mu}(z)=k_i(n_i^{\mathrm{CR}}-n_i^{t})G(z)
\end{equation}
In the above equation, $k_i$ is a force constant and $n_i^{t}$ is the target number density of species $i$ in the CRs.
The force is modelled according to a continuous function centered at $z_F$:
\begin{equation}
    G(z)=\frac{1}{4 \omega} \left[ 1 + \mathrm{cosh} \left( \frac{z_0-z_F}{\omega} \right) \right]^{-1}
    \label{eqn:bellfunction}
\end{equation}
The function can be made arbitrarily sharp by changing $\omega$. 
In the simulations in this work, $\omega$ was 0.001\% of the total extent of $z$; $z_F$ was 5.9~nm and $z_0$ was the $z$ position of the centre of the solid slab; the size of the CRs in $z$ was 2.2~nm; and, $k_i=2 \times 10^5$~kJ~mol$^{-1}$. 
Simulations were performed for at least 100~ns with target ion number densities in the range $n^t=$ 0.6022--6.022, corresponding to ion molar concentrations, $c^t=n^t \times 10^{24}/N_{\mathrm{A}}$ (where $N_{\mathrm{A}}$ is Avogadro's constant), of 1--10 M (mol dm$^{-3}$) in the case of simulations including graphite and $c^t=1,3,5~\mathrm{and}~7$~M otherwise. 

\subsection{Preparation of initial configurations.}
\noindent A $5.4 \times 5.5 \times 2.7$~nm ($x \times y \times z$) graphite supercell\cite{trucano_structure_1975} was constructed; this comprised eight graphene layers and was positioned in the centre of an orthorhombic simulation cell such that the $c$-axis was parallel to $z$.
The graphite was placed in contact with an aqueous solution containing 1,672 NaCl and 13,819 water molecules (for a total NaCl molality of 6.7~mol/kg).
With restraints applied to the position of carbon atoms to avoid distortion of the solid, a molecular dynamics (MD) simulation was performed for 0.2~ns at 298~K and 1~bar to relax the solution and equilibrate the simulation cell volume.
The pressure of the system was held constant within statistical fluctuations using the barostat of Berendsen et al. \cite{berendsen_molecular_1984} 
With the volume held constant, a series of simulations were subsequently performed to prepare the ion-rich reservoir.
Here, harmonic restraints were applied to the distances between ions in solution and a point at the centre of the simulation cell (where the solid slab was positioned).
In a series of 0.2--0.5~ns simulations, the minimum energy distance for the restraints was increased from 1 to 6~nm with a force constant of $3 \times 10^5$~kJ~mol$^{-1}$.

The protocol to prepare the initial configuration for simulations for NaCl(s) in contact with NaCl(aq) was the same as in the case of graphite; however, here a $5.6 \times 5.6 \times 2.8$~nm NaCl rock-salt supercell formed the solid substrate.
To prepare `rough' crystal surfaces, 79 NaCl ion pairs were removed from the uppermost layers of one side of the {100} NaCl(s) surface and the structure, in contact with water, was then optimised before performing a 0.2~ns simulation to relax the surface further.
NaCl solids were also restrained in the simulation cell by applying restraints to eight ions at the centre of the slab to their initial positions using harmonic potential biases with $k=600$~kJ~mol$^{-1}$.

\subsection{Self-Diffusion calculations.}
\noindent The self-diffusion coefficients ($D_{self}$) for ions and water were measured in simulations of bulk solutions which approached infinite dilution. Single ions were placed into bulk liquid water comprising 4,000 water molecules. 
Simulations were performed at 298~K and 1~bar using the thermostat described above and the Parinello-Rahman barostat, \cite{parrinello_polymorphic_1981} respectively.
50~ns simulations were performed using a 1~fs timestep, from which the mean squared displacement of atoms was calculated according to
\begin{equation}
    \mathrm{MSD} = \frac{1}{N} \sum_{i=1}^N \langle | \mathbf{r}(t)-\mathbf{r}(0) |^2\rangle
\end{equation}
where the time origin was reset every 10~ps. A linear fit to $\mathrm{MSD}(t)$ in the range $1-10$~ns allowed for the calculation of the apparent self-diffusion coefficient ($D$) according to $\mathrm{MSD}(t)=6Dt$. 
 A correction was applied to account for the finite system size, \cite{yeh_system-size_2004} resulting in $D_{self}$ for Na$^+$, Cl$^-$ and H$_2$O of $1.223 (0.005), 1.282 (0.008)~\mathrm{and}~2.762 (0.021) \times 10^{-5}~\mathrm{cm}^2 ~\mathrm{s}^{-1}$, respectively. 

\subsection{The classical force field.}
\noindent Pairwise intermolecular interactions between ions and extended simple pint charge (SPC/E) water molecules\cite{berendsen_missing_1987} were modelled using the Joung and Cheatham force field, \cite{joung_determination_2008}
The equilibrium solute molality evaluated for this force field is 3.7~mol/kg. \cite{benavides_consensus_2016}
Graphite C--C atomic interactions were modelled using the OPLS/AA force field; \cite{jorgensen_development_1996} C--water intermolecular interactions were modelled using the atom pair potential provided by Wu and Aluru, \cite{wu_graphitic_2013}, which was based on fitting to water adsorption energies from random phase approximation calculations;\cite{ma_adsorption_2011} while C--ion intermolecular interactions were modelled using the atom pair potentials from Williams et al., \cite{williams_effective_2017} based on density functional theory calculations of the adsoprtion of ions in a continuum polarisable cavity.
Both the C--water and C--ion potentials were fitted to the same three point charge model for water, ensuring a self-consistency in the force field used here.

\subsection{Data availability} GROMACS and Plumed input and example output files, including the force field parameters necessary to reproduce the simulation results reported in this paper, are available on github (https://github.com/aaronrfinney/CmuMD-NaCl\_at\_graphite). The PLUMED input files are also accessible via PLUMED-NEST (https://www.plumed-nest.org\cite{bonomi2019promoting}), the public repository for the PLUMED consortium, using the project IDs: plumID:21.035 (NaCl(s) surfaces) and plumID:21.011 (Graphite surfaces). Details on how to use and implement the C$\mu$MD method within PLUMED is available on github (see https://github.com/mme-ucl/CmuMD).

\subsection{Acknowledegments}
The authors acknowledge funding from an EPSRC Programme Grant (Grant EP/R018820/1). The authors acknowledge the use of the UCL Myriad High Throughput Computing Facility (Myriad@UCL), and associated support services, in the completion of this work.

\section*{References}
\bibliographystyle{achemso}
\bibliography{int}

\end{document}